\documentclass[aps,prc,superscriptaddress,twoside,twocolumn,%
nofootinbib,showpacs]{revtex4}
\usepackage{amsmath,amssymb}
\usepackage{graphicx}

\begin{document}

\title{\boldmath Interaction current in $p p \to p p \gamma$}

\author{K. Nakayama}
\email{nakayama@uga.edu}

\affiliation{Department of Physics and Astronomy, University of Georgia,
Athens, Georgia 30602, USA} \affiliation{\mbox{Institut f{\"u}r Kernphysik and
J\"ulich Center for Hadron Physics, Forschungszentrum J{\"u}lich, 52425
J{\"u}lich, Germany}}

\author{H. Haberzettl}
\email{helmut.haberzettl@gwu.edu}

\affiliation{\mbox{Center for Nuclear Studies, Department of Physics, The
George Washington University, Washington, DC 20052, USA}}

\date{Received 2 October, 2009; published 11 November, 2009}

\begin{abstract}
The nucleon-nucleon bremsstrahlung reaction is investigated based on a fully
gauge-invariant relativistic meson-exchange model approach. To account
consistently for the complicated part of the interaction current (which at
present is too demanding to be calculated explicitly), a generalized contact
current is introduced following the approach of H. Haberzettl, K. Nakayama, and
S. Krewald [Phys.\ Rev.\ C\,\textbf{74}, 045202 (2006)]. The contact
interaction current is constructed phenomenologically such that the resulting
full bremsstrahlung amplitude satisfies the generalized Ward-Takahashi
identity. The formalism is applied to describe the high-precision proton-proton
bremsstrahlung data at 190 MeV obtained at KVI [H. Huisman \textit{et al.},
Phys.\ Rev.\ C\,\textbf{65}, 031001(R) (2002)]. The present results show good
agreement with the data, thus removing the long-standing discrepancy between
the theoretical predictions and experimental data. The present investigation,
therefore, points to the importance of properly taking into account the
interaction current for this reaction.
\end{abstract}

\pacs{25.10.+s,
      25.40.-h,
      25.20.Lj,
      13.75.Cs
      \hfill[Phys.~Rev.~C\,\textbf{80}, 051001(R) (2009)]}

\maketitle


The nucleon-nucleon ($NN$) bremsstrahlung reaction had been studied extensively
in the past mainly to learn about off-shell properties of the $NN$ interaction.
It should be clear, however, that off-shell effects are model-dependent and
therefore are meaningless quantities for comparison. In fact, Fearing and
Scherer \cite{FS00} have shown explicitly for $NN$ bremsstrahlung and related
processes that in field theories off-shell effects cannot be measured.

\begin{figure*}[t!]\centering
\includegraphics[width=\textwidth]{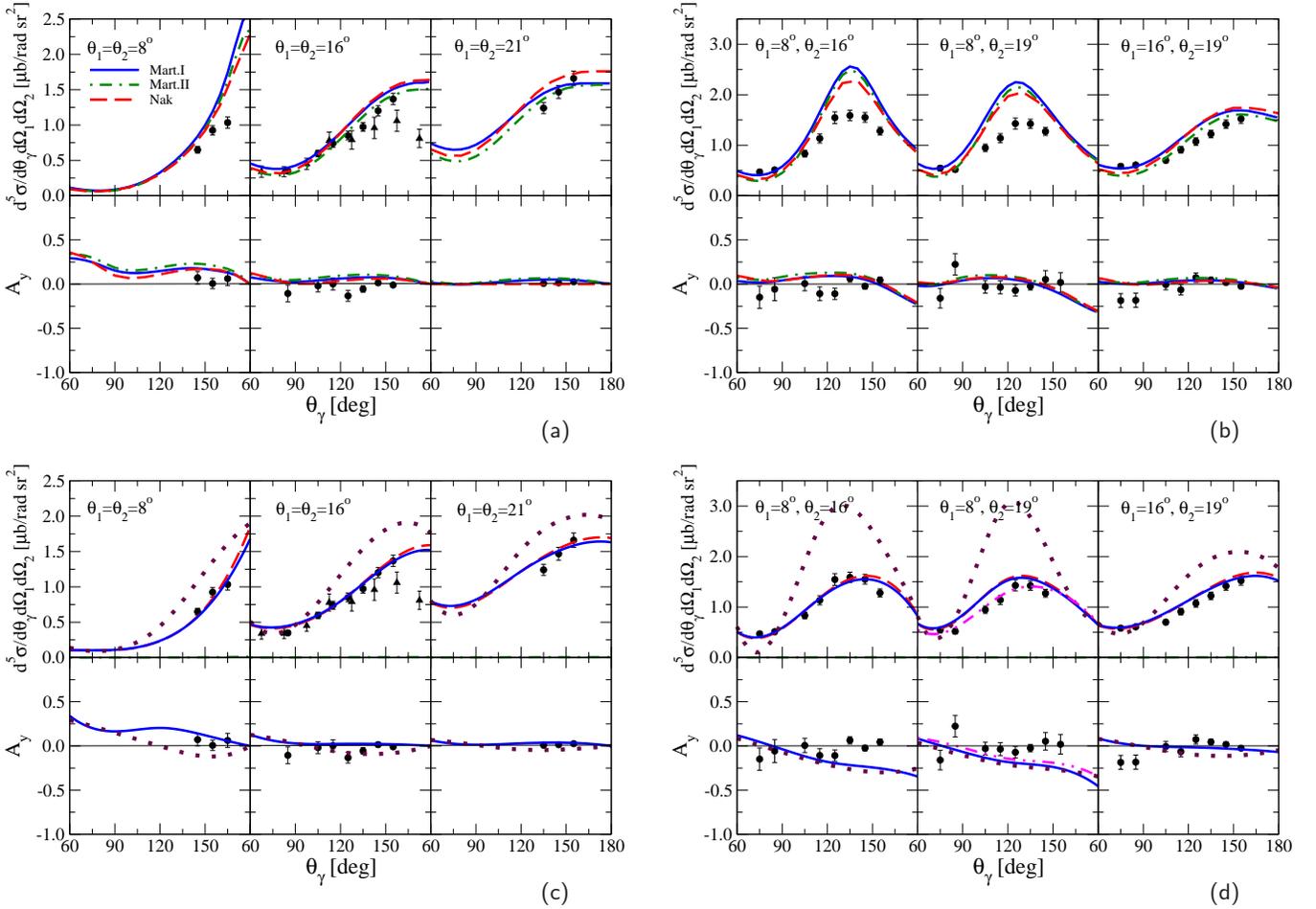}
\caption{\label{fig:1}%
(Color online) Comparison of the KVI data \cite{KVI02} (with corrected $A_y$
sign~\cite{KVI04}) for  $pp \to pp\gamma$ in coplanar geometry at 190 MeV
proton incident energy (circles) with various theoretical predictions. The
TRIUMF data \cite{TRIUMF80} for cross sections (triangles) are also displayed
here. In each figure, the upper rows of panels display the cross sections and
the lower rows the corresponding analyzing powers $A_y$ for fixed proton
scattering angles, $\theta_1$ and $\theta_2$, and as functions of the emitted
photon angle, $\theta_\gamma$, in the laboratory frame. The two figures on the
left are for the symmetric proton scattering angles, $\theta_1=\theta_2$, while
the two figures on the right are for the asymmetric proton angles, $\theta_1
\ne \theta_2$.
Figures (a) and (b) show the \emph{results of previous model calculations}. The
solid lines are the results of Ref.~\cite{MST97} including all the higher-order
corrections; the dash-dotted lines pertain to the same model of
Ref.~\cite{MST97} without the higher-order corrections. The dashed lines
represent the results of Ref.~\cite{HNSA95}.
Figures (c) and (d) show the same data arranged the same way compared to
\emph{results obtained with the present model}, i.e., the theoretical results
shown in (c) are to be compared to those of (a) and those of (d) to (b). The
dashed curves in (c) and (d) represent the nucleonic plus generalized contact
current contributions; the dash-dotted curves correspond to the mesonic
current, while the solid curves denote the total current contribution.
 Switching off the generalized four-point contact
current (\ref{eq:2}) produces the dotted curves. The dashed-double-dotted lines
for the middle panels in (d) exhibit the parameter sensitivity explained in the
text. } \vspace{-1.8mm}
\end{figure*}

Even though the original motivation for investigating the $NN$ bremsstrahlung
reaction has fallen away, understanding the dynamics of the $NN$ bremsstrahlung
reaction, nevertheless, is of extreme importance in general for it is one of
the most fundamental processes involving both electromagnetic and hadronic
interactions. Its importance is all the more emphasized by the fact that, so
far, none of the existing models of $NN$ bremsstrahlung can describe the
high-precision proton-proton bremsstrahlung data from KVI \cite{KVI02,KVI04}
for coplanar geometries involving small proton scattering angles. This is
illustrated in Figs.~\ref{fig:1}(a) and (b), where the KVI data \cite{KVI02}
for cross sections and analyzing powers are compared to the results of the
microscopic calculations of Martinus \textit{et al.} \cite{MST97} and of
Herrmann \textit{et al.} \cite{HNSA95}. Also shown in Fig.~\ref{fig:1} are the
TRIUMF data for the cross section~\cite{TRIUMF80}. As one can see, all the
model calculations overestimate the measured cross sections, especially, for
asymmetric proton scattering angles ($\theta_1 \ne \theta_2$). In
Ref.~\cite{KVI02}, some soft-photon model results are also compared with the
data. In contrast to the microscopic models, these soft-photon models reproduce
well the measured cross-section data. For the analyzing powers, however, it is
the microscopic models that describe the data much better than the soft-photon
models. (See also Ref.~\cite{KVI09}, where new data from KVI were reported
sampling a part of phase space different from that for the earlier data
\cite{KVI02,KVI04}; the soft-photon models are found to be at odds with these
new data.) It should be noted here that, strictly speaking, the kinematical
regime of the KVI data for small proton scattering angles are outside of the
range of applicability of Low's soft-photon theorem~\cite{L58}. There are also
a number of other microscopic model calculations available in the literature
\cite{KA93,J94,JF95,EG96,KMS98,CMSTT02,CST03}, which are dynamically similar to
Refs.~\cite{MST97,HNSA95}, addressing a variety of issues in the $pp$
bremsstrahlung process.

A detailed discussion of the status of the discrepancy between the theoretical
and the experimental results in $pp$ bremsstrahlung can be found in
Refs.~\cite{CMSTT02,CST03,KVI02}. This situation is extremely disturbing from a
theoretical point of view, in particular, if one considers the fact that these
data are obtained at a proton incident energy of only 190 MeV, well below the
pion-production threshold energy of about 280 MeV. At such a low energy, one
expects the nucleonic current to be dominating by far, and baryon resonances as
well as meson-exchange currents should play only minimal roles in the reaction
dynamics.

%
%

In this work we show that the interaction current mandated by gauge invariance
plays a crucial role for this reaction and that its proper inclusion in
theoretical models removes, to a large extent, the existing discrepancy between
the theoretical and the experimental results. For direct and easy comparison,
our present numerical results --- explained later in this article --- are shown
in Figs.~\ref{fig:1}(c) and (d), directly below those shown in
Figs.~\ref{fig:1}(a) and (b) obtained by other authors \cite{MST97,HNSA95}.

The present work uses a novel approach to describe the $NN$ bremsstrahlung
reaction. It is derived within a relativistic field-theory approach by coupling
the photon everywhere possible in the underlying two-nucleon $T$-matrix
determined by the corresponding $NN$ Bethe-Salpeter equation. The basic idea of
this formalism is the same as what had been introduced by Haberzettl, Nakayama,
and Krewald~\cite{HNK06} for pion photoproduction, based on the
field-theoretical approach of Haberzettl \cite{H97}.

%
\begin{figure*}[t!]\centering
\includegraphics[width=\textwidth,clip=]{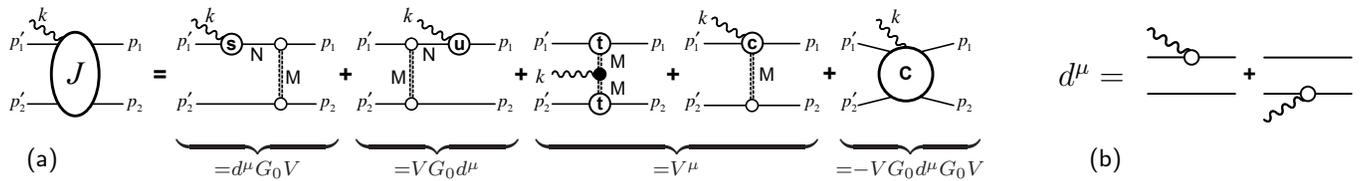}
\caption{\label{fig:2}%
(a) Basic production amplitude $J^\mu$ of eq.~(\ref{eq:1J}) for $NN\to NN
\gamma$ used in the present work. (Time proceeds from right to left.) The terms
below the diagrams correspond to the respective ones in Eq.~(\ref{eq:1J}), with
(b) showing the photon coupling to both intermediate nucleons subsumed in
$d^\mu$.  The diagrams for the lower nucleon line analogous to diagrams 1, 2,
and 4 are suppressed. $N$ denotes the intermediate nucleon and $M$ incorporates
all exchanges of mesons $\pi$, $\eta$, $\rho$, $\omega$, $\sigma$, and $a_0^{}$
(former $\delta$). In general, contributions to the $NN$ interaction more
complex than single-meson exchange may be considered as well. External legs are
labeled by the four-momenta of the respective particles; the hadronic vertices
$s$, $u$, and $t$ (with the labels alluding to the corresponding Mandelstam
variables) correspond to the same kinematic situations, respectively. The first
two diagrams on the right-hand side describe the so-called nucleonic current
and the meson-exchange current is depicted by the third diagram. The fourth
diagram contains the $N M\to N \gamma$ four-point contact current of
Eq.~(\ref{eq:2}), labeled ``c" in the diagram. The first four diagrams
correspond to the \emph{complete} gauge-invariant description for the process
$NM\to N\gamma$ for the upper nucleon line. The last diagram (labeled ``C")
stands for the five-point contact-type current in general necessary to preserve
gauge invariance of the entire amplitude, including the $NN$ ISI and FSI
contributions.}
\end{figure*}
%
%

\enlargethispage{2pt}%
The full bremsstrahlung amplitude can be written as
\begin{equation}
M^\mu = (TG_0+1)J^\mu(1+G_0T) \ ,
\label{eq:1}
\end{equation}
where the $NN$ $T$-matrices on the left and right mediate the final-state (FSI)
interaction and the initial-state interaction (ISI), respectively; $G_0$
denotes the intermediate propagation of two free nucleons. The current
\begin{equation}
J^\mu =d^\mu G_0 V +VG_0 d^\mu +V^\mu - VG_0d^\mu G_0 V
\label{eq:1J}
\end{equation}
is the basic photon production current off the two nucleons composed, in
general, of nucleonic, mesonic, and baryon-resonance currents, in addition to
contact-type interaction currents, as shown in Fig.~\ref{fig:2}(a) for the
example of single-meson exchanges. The two disconnected nucleonic contributions
subsumed in $d^\mu$ are shown in Fig.~\ref{fig:2}(b); $V$ is the $NN$
interaction and $V^\mu$ describes the photon coupling to the internal
mechanisms of the interaction $V$. We emphasize that the generic structures of
Eqs.~(\ref{eq:1}) and (\ref{eq:1J}) are exact for any type of $NN$ interaction.
The proof follows straightforwardly from applying the ``gauge derivative"
procedure given in the Appendix of Ref.~\cite{H97} to the connected part
$G_0TG_0$ of the $NN$ Green's function. For $NN$ interactions based on
single-meson exchanges, the structure of $J^\mu$ depicted in Fig.~\ref{fig:2}
is complete. We draw particular attention to the contact-type currents
appearing in the fourth and fifth term in Fig.~\ref{fig:2}(a). In general, they
possess very complex internal dynamical structures that, at present, cannot be
taken into account explicitly. However, they are constrained by gauge
invariance. The corresponding constraint for the four-point current in the
fourth diagram is based on the generalized
Ward-Takahashi identity for the subprocess $NM\to N\gamma$.\footnote{\label{foot1}%
See the corresponding discussion in Ref.~\cite{HNK06} for the case of pion
photoproduction that exemplifies the structure of the four-point contact
current appearing in the fourth diagram of Fig.~\ref{fig:2}(a). This case is
relevant here because the dynamics of $NN$ bremsstrahlung can be largely
understood as a meson capture process where the captured meson originates from
a spectator nucleon, as can be seen in the first four diagrams in
Fig.~\ref{fig:2}(a).} By contrast, the five-point contact current depicted in
the last diagram of Fig.~\ref{fig:2}(a) is constrained by demanding gauge
invariance of the full bremsstrahlung process using the fact that the
subprocesses already satisfy their respective gauge-invariance constraints.

For the current $J^\mu$ considered here shown in Fig.~\ref{fig:2}(a), in
addition to the first two diagrams on the right-hand side with intermediate
nucleons marked $N$, one may also consider contributions from intermediate
baryon resonances. However, for the present application to the KVI data
\cite{KVI02} at 190\,MeV incident proton energy, we expect their contributions
to be minimal, and we therefore have omitted such contributions.\footnote{For
the $\Delta$, in particular, we point to the results of  Ref.~\cite{MST97}
shown in Figs.~\ref{fig:1}(a) and (b) with (solid curves) and without
(dashed-dotted curves) higher-order effects, where the former includes the
$\Delta$ as well as the mesonic current contributions. They show that the
$\Delta$ would have only a minor effect in the low-energy regime investigated
here. This therefore cannot affect our overall conclusions. At higher energies,
effects of the $\Delta$ resonance have been investigated by de Jong \textit{et
al.} \cite{J94}.}

Our present approach is fully relativistic, employing a covariant
three-dimensional reduction of the Bethe-Salpeter equation underlying the $NN$
interaction~\cite{Bonn_NN}. As an immediate consequence of this reduction, one
finds that the five-point contact current [last diagram in Fig.~\ref{fig:2}(a)]
must not contribute to the four-divergence of the full amplitude, i.e., it must
be fully transverse. (This is \emph{not} true for the amplitude $M^\mu$
evaluated in a full four-dimensional framework. The five-point contact current
then is \emph{essential} to maintaining gauge invariance.) In our present
calculation, therefore, because it has no bearing on gauge invariance, we have
dropped the five-point contact current entirely. Further details of the present
formalism will be reported elsewhere.

In view of this finding, gauge-invariance constraints come to bear only on the
four-point interaction current describing the $NM \to N \gamma$ subprocess in
the fourth diagram of Fig.~\ref{fig:2}(a). To avoid having to deal with its
very complex microscopic dynamical structures~\cite{HNK06}, we employ a
generalized contact term that is constructed such that the resulting full
amplitude satisfies the generalized Ward-Takahashi identity necessary to
ensures full gauge invariance. The details of the contact term employed here
are discussed later in this article in connection with Eq.~(\ref{eq:2}). The
only mesonic currents [cf.\ third diagram in Fig.~\ref{fig:2}(a)] contributing
to $pp$ bremsstrahlung are those arising from the anomalous couplings which
cannot be obtained from coupling the photon to the underlying $NN$ interaction.
Following Ref.~\cite{HNK06}, we include the $\pi\rho\gamma$- and
$\pi\omega\gamma$-exchange contributions in the present work which are the
dominant mesonic currents for $pp$ bremsstrahlung.

In the present work, we use the OBEP-B version of the Bonn $NN$ interaction
with its parameters slightly readjusted to reproduce the low-energy $pp$
scattering length \cite{Bonn_NN} (see also Ref.~\cite{HNSA95}). This
interaction contains only nucleon and meson degrees of freedom. The reason for
choosing this interaction is that, apart from the phenomenological form factors
at the nucleon-nucleon-meson ($NNM$) vertices, it is a fully microscopic
meson-exchange model and, as such, the photon can be attached consistently and
uniquely to every part of the $NN$ interaction except, of course, to the form
factors. The latter mechanism introduces necessarily an ambiguity in how the
photon couples to this interacting system. As alluded to previously, we account
for it through the generalized four-point contact current given below. We note
once more that with such a simple explicit microscopic model of the $NN$
interaction based on a three-dimensional reduction of the underlying
Bethe-Salpeter equation, no five-point contact current is required to maintain
gauge invariance, i.e., all currents can be calculated explicitly, with the
exception of the four-point contact current arising from the photon coupling to
the $NNM$ vertices, as pointed out. A five-point contact current may be
necessary when one uses a more sophisticated $NN$ interaction that includes
two-nucleon irreducible contributions, as would be the case, for example, with
interactions including explicit $\Delta$ degrees. For such interactions, and
also for those based on purely phenomenological approaches where the underlying
microscopic structures are not known, the introduction of a five-point contact
current becomes an unavoidable procedure.

As alluded to in footnote~\ref{foot1}, it is instructive to consider the
present approach from a different point of view. Ignoring for the moment the
complication of a possible five-point contact current, the basic photon
production amplitude $J^\mu$ can be viewed as being obtained from the inverse
of the photoproduction reaction, i.e., a meson capture reaction, where the
captured meson originates from a second nucleon that is a spectator to the
capture reaction. In this picture, the nucleon-meson FSI is accounted for
effectively by the four-point contact current. This is discussed in detail in
Ref.~\cite{HNK06}. In this sense, one may think of the full bremsstrahlung
amplitude given by Eq.~(\ref{eq:1}) as being constructed from the two basic
building blocks, the respective amplitudes for the $NN \to NN$ and the $NM \to
N\gamma$ reactions, analogous to the approach described in Ref.~\cite{NSL02}
for calculating the $NN \to NN\eta$ process. Of course, in this simplified
picture one ignores complications arising from the identity of the nucleons.

\enlargethispage{3pt} For the details of the derivation of the generalized
four-point contact current involving a pseudoscalar meson exchange, we refer to
Ref.~\cite{HNK06}. Here, the required extension --- to account for the virtual
nature of the incoming and outgoing nucleons and the exchanged meson at the
four-point vertex
--- is accomplished following the work of Ref.~\cite{NOH06}. Furthermore, in
the present work, the generalized four-point contact current is extended to
include also the scalar and vector meson exchanges. Schematically, suppressing
all Lorentz indices, we have for the fourth diagram in Fig.~\ref{fig:2}(a)
\begin{equation}
J^\mu_{c_1} = \sum_M \left(f_{2t}\Gamma_2\right) i\Delta_M
               \left[e_M f_{1t} \Gamma^\mu_{c_1} + \Gamma_1 \tilde{C}^\mu_1 \right]~,
\label{eq:2}
\end{equation}
where the summation runs over the exchanged mesons between the interacting
nucleons 1 and 2. Nucleon 2 (described by its phenomenological form factor
$f_{2t}$ and its Lorentz operator structure $\Gamma_2$) is here the spectator
nucleon that supplies the meson (described by the propagator $\Delta_M$)
participating in the $NM\to N\gamma$ subprocess at the other vertex. The
corresponding four-point interaction current is described by the expression in
the square brackets here. The first term contains the phenomenological form
factor $f_{1t}$ and current operator $\Gamma^\mu_{c_1}$, i.e., this describes
the usual Kroll-Ruderman-type $NNM\gamma$ contact vertex. The second term
contains the Lorentz operator $\Gamma_1$ for the $NNM$ vertex of nucleon 1 and
the contact current~\cite{HNK06,H97}
\begin{align}
\tilde{C}^\mu_1 \equiv & - e'_1 \frac{(2p'_1 + k)^\mu}{(p'_1+k)^2 - {p'_1}^2}(f_{1s}-\hat F_1)
\nonumber\displaybreak[2] \\
                &        -  e_1 \frac{(2p_1 - k)^\mu}{(p_1-k)^2 - p_1^2}(f_{1u}-\hat F_1)
\nonumber\displaybreak[2] \\
                &        -  e_M \frac{(2q - k)^\mu}{(q-k)^2 - q^2}(f_{1t}-\hat F_1)
\label{eq:3}
\end{align}%
that is necessary for the preservation of gauge invariance because the nucleons
have structure described by form factors. The factors $e'_1$, $e_1$, and $e_M$
stand for the combined charge-isospin operators of the nucleons and meson at
the four-point vertex. The subtractions of $\hat{F}_1$ in Eq.~(\ref{eq:3}) are
necessary to render $\tilde{C}^\mu_1$ pole-free; $\hat{F}_1$ is a
phenomenological function chosen here as
\begin{equation}
\hat F_1 = R_1
- h\frac{(R_1 - \delta_s f_{1s})(R_1 - \delta_u f_{1u})(R_1 - \delta_t f_{1t})}{R_1^2}~,
\label{eq:4}
\end{equation}
which is manifestly crossing symmetric. For nonzero charges $e_x$, one has
$\delta_x=1$ and zero otherwise; $f_{1x}$ denote the hadronic form factor for
the specified kinematics $x=s,u,t$ [cf.\ Fig.~\ref{fig:2}(a)]; $p'_1$ and $p_1$
are the four-momenta at the four-point vertex of the outgoing and incoming
nucleon, respectively; and $q$ and $k$ stand for the four-momenta of the
exchanged meson $M$ and the emitted photon, respectively. In Eq.~(\ref{eq:4}),
\begin{equation}
R_1 = 1 + e^{-z/a}(f - 1) \ ,
\label{eq:5}
\end{equation}
with $z = \big[ (1 - \delta_s f_{1s})(1 - \delta_u f_{1u})(1 - \delta_t f_{1t})
\big]^2$, and $f=F({p'_1}^2, p_1^2, q^2)$ denotes the hadronic form factor with
the momentum arguments as indicated. Of course, an expression similar to
Eq.~(\ref{eq:2}) exists for the photon emerging from the $NNM$ vertex of
nucleon 2. [The latter diagram is not shown in Fig.~\ref{fig:2}(a).]

The only free parameters of our model are the parameters $h$ and $a$ appearing
in Eqs.~(\ref{eq:4}) and (\ref{eq:5}) in the generalized four-point contact
currents involving the scalar, pseudoscalar, and vector meson exchanges [cf.\
Fig.~\ref{fig:2}(a), fourth diagram]. In principle, these parameters may both
be chosen independently for different exchanged mesons, in addition to being
functions of momenta at the four-point vertex. In the present work, we take
them to be constant and equal for all the exchanged mesons for the sake of
simplicity. Their values of $h=2.5$ and $a=1000$ have been adjusted to
reproduce the KVI cross section data \cite{KVI02}. All coupling constants and
form factors at the hadronic vertices are consistent with the Bonn $NN$
interaction \cite{Bonn_NN} we use here for the FSI and the ISI. The only
exception is the $NN\omega$ coupling constant, $g_{NN\omega}$, and the cutoff
parameter, $\Lambda_\pi$, in the $NN\pi$ form factor entering in the mesonic
current. Following the discussion in Ref.~\cite{NSHHS98}, we take
$g_{NN\omega}=10$ and $\Lambda_\pi=900$ MeV. Also, following the work of
Refs.~\cite{NSHHS98,NSL02}, we employ a form factor for the off-shell nucleon
in the basic photon production current, $J^\mu$, with a cutoff parameter value
of $\Lambda_N=1000$ MeV.

In Figs.~\ref{fig:1}(c) and \ref{fig:1}(d), we show the present results for the
cross sections and analyzing power for the symmetric (left panels) and
asymmetric (right panels) proton scattering angles. As one can see, we now
reproduce well the cross-section data for both symmetric and asymmetric proton
scattering angles (solid curves). The dotted curves correspond to the results
when the generalized contact current is switched off. This illustrates the
importance of taking into account the interaction current properly, a feature
that has been ignored in all the earlier models. In fact, our results without
the contact current are very similar to the ones obtained in earlier models,
especially for asymmetric proton scattering angles [compare, in particular, the
top-row panels in Figs.~\ref{fig:1}(a) and \ref{fig:1}(b) with those in
Figs.~\ref{fig:1}(c) and \ref{fig:1}(d), respectively]. The contact current,
however, does not affect the analyzing power significantly and thus leaves room
for further improvements.

To provide some idea about how sensitive the present results are to the fit
parameters $h$ and $a$, the dashed-double-dotted curves in the middle panels of
Fig.~\ref{fig:1}(d), for $\theta_1=8^\circ$, $\theta_2=19^\circ$, correspond to
$h=3.0$ in Eq.~(\ref{eq:4}). Note that the agreement with the data improves for
both the cross section and the analyzing power. The calculated results are
rather insensitive to the parameter $a$ in Eq.~(\ref{eq:5}), once it is in the
correct order-of-magnitude range.

The Bonn $NN$ interaction \cite{Bonn_NN} used here does not incorporate Coulomb
effects. To account for them, we repeated the calculation using instead the
Paris $NN$ interaction \cite{Paris_NN}, which includes the Coulomb interaction
fully as described in Ref.~\cite{HNSA95}. We obtain results practically the
same as the ones shown in Figs.~\ref{fig:1}(c) and \ref{fig:1}(d) if we
readjust the parameter value $h$ in Eq.~(\ref{eq:4}) to $h=2.0$. The only
noticeable Coulomb effect is a downward bending of the cross section near
$\theta_\gamma = 180^\circ$ in the $\theta_1=\theta_2=8^\circ$ geometry, which
is in agreement with earlier findings \cite{KA93,HNSA95}.

In summary, our present results essentially resolve the long-standing
discrepancy between theoretical and experimental results in the $pp$
bremsstrahlung reaction. The new feature of the present model responsible for
bringing the theoretical results in line with the measured high-precision
cross-section data from KVI \cite{KVI02} is a generalized four-point contact
current that accounts for the interaction current in the $NM \to N \gamma$
subprocess and that is constructed in such a way that the resulting full
bremsstrahlung amplitude obeys the Ward-Takahashi identity and thus is gauge
invariant.

In view of its phenomenological nature we do not expect the particular contact
current employed here to provide a definitive resolution of all problems in
$NN$ bremsstrahlung processes. Nevertheless, our present results show that the
interaction current is a necessary ingredient for bremsstrahlung calculations
that cannot be neglected.

This work is supported in part by the FFE-COSY Grant
No.\ 41788390.


\end{document}